\documentclass[aps]{revtex4}
\usepackage{multirow}
\usepackage{float}
\usepackage{color,graphicx,amsmath,amssymb,bbold}
\newcommand{\be}{\begin{equation}}
\newcommand{\ee}{\end{equation}}
\begin{document}
 \title{Superharmonic  double-well  systems with  zero-energy ground states: Relevance for diffusive  relaxation scenarios}
 \author{Piotr Garbaczewski and Vladimir A. Stephanovich}
 \affiliation{Institute of Physics, University of Opole, 45-052 Opole, Poland}
 \date{\today }
 \begin{abstract}
 Relaxation properties (specifically time-rates)  of the  Smoluchowski  diffusion process on a line, in a confining  potential  $ U(x) \sim    x^m$,   $m=2n \geq 2$,  can be spectrally  quantified by means of the affiliated  Schr\"{o}dinger semigroup $\exp (-t\hat{H})$, $t\geq 0$. The inferred  (dimensionally rescaled)  motion generator   $\hat{H}= - \Delta + {\cal{V}}(x)$  involves a potential function  ${\cal{V}}(x)= ax^{2m-2} - bx^{m-2}$,  $a=a(m),  b=b(m) >0$, which for $m>2$   has a conspicuous  higher  degree (superharmonic) double-well  form. For each value of  $m>2$,  $ \hat{H}$ has   the   zero-energy  ground state eigenfunction   $\rho _*^{1/2}(x)$,   where  $\rho _*(x) \sim \exp -[U(x)]$  stands for the  Boltzmann equilibrium pdf of  the  diffusion process.  A  peculiarity of   $\hat{H}$  is that it  refers to a family of   quasi-exactly solvable  Schr\"{o}dinger-type systems,  whose  spectral data  are either  residual  or analytically  unavailable. As well,  no numerically assisted procedures have been developed to  this  end.  Except for the  ground state zero eigenvalue  and  incidental  trial-error outcomes,  lowest positive  energy levels (and energy  gaps)  of  $\hat{H}$  are unknown.   To overcome this  obstacle, we develop  a computer-assisted procedure  to recover an  approximate spectral solution  of $\hat{H}$ for  $m>2$.  This  task  is  accomplished  for the relaxation-relevant  low part of the   spectrum.  By admitting larger values of $m$ (up to   $m=104$), we   examine the  spectral "closeness"   of  $\hat{H}$, $m\gg 2$ on $R$   and   the   Neumann Laplacian $\Delta _{\cal{N}}$ in the  interval $[-1,1]$,  known  to  generate  the    Brownian motion with two-sided reflection.
 \end{abstract}
 \maketitle

\section{Introduction.}

In the presence of confining conservative forces,  the Smoluchowski (Fokker-Planck) equation, here considered in one dimension,   $\partial _t \rho  = D \Delta \rho - \nabla (b \rho )= L^*\rho $, takes over an initial probability density function (pdf)  $\rho _0(x)$  to an  asymptotic stationary ($L^*\rho _*(x)=0$)  pdf of the Boltzmann form $\rho _*(x)= (1/Z) \exp[- U(x)/D]$.  Here  $b(x)= - \nabla U(x)$ and $Z$ is the   $L^1(R)$-normalization constant,  $D$   stands for  the diffusion coefficient, which   upon suitable rescaling   may be set equal $1$ (the value  $1/2$ is often employed in the mathematically oriented research).  For the record,  we mention that  $L^*$ denotes the Fokker-Planck operator, while $L$  the diffusion generator of the stochastic process,  \cite{jph,pavl}.

 Given a stationary pdf $\rho _*(x)$, one can  transform the $L^1(R)$  Smoluchowski-Fokker-Planck evolution  $\exp(tL_*)$, $t\geq 0$ to the  $L^2(R)$   Schr\"{o}dinger semigroup  $\exp(-t \hat{H})$, see e.g.   \cite{jph}-\cite{jph1}.  A classic factorisation  \cite{risken}  of $\rho (x,t)= \Psi (x,t) \rho _*^{1/2}(x)$  allows to map  the Fokker-Planck  dynamics into the  the generalized  (Schr\"{o}dinger-type)  diffusion problem. In the dimensionally  rescaled ($D=1$) form we have:
   \be
   \partial _t \Psi =  \Delta  \Psi - {\cal{V}} \Psi = - \hat{H} \Psi .
   \ee
Accordingly,  the   relaxation process $\rho (x,t) \to \rho _*(x)$ is paralleled by  the $L^2(R)$   relaxation    $\Psi _0(x) \to \Psi (x,t) \to \rho _*^{1/2}(x)$.
  We have  $\hat{H} \rho _*^{1/2}= 0$, hence   $\rho _*^{1/2}$ is a legitimate {\it zero  energy} bound state of $\hat{H}$. Moreover, the functional form of the induced (Feynman-Kac by provenience, \cite{jph,vilela,faris})  potential ${\cal{V}}(x)$  readily  follows:
\be
{\cal{V}}(x) =  {\frac{\Delta \rho _*^{1/2}}{\rho _*^{1/2}}} =     \frac{1}{2} \left(\frac{b^2}{2} + \nabla b\right),
\ee
We note that  $b(x)=  \nabla \ln \rho _*(x)= -\nabla U(x)$, \cite{jph},  \cite{streit}-\cite{turbiner1}.\\

 The eigenvalues of $\hat{H}$, up to an inverted  sign,  are shared by the Fokker-Planck operator    $L^* = \Delta - \nabla [b(x)\, \cdot ] $ and  the diffusion generator  $L = \Delta + b(x) \nabla $,  \cite{jph,pavl}.
 If we have the spectral solution for   $\hat{H}$ in hands, in terms of eigenvalues $\lambda$ and $L^2(R)$  eigenfunctions $\Psi _{\lambda }(x)$,  then the eigenvalues of $L^*$  are  $- \lambda$,  while the  corresponding  eigenfunctions appear in the form    $\phi _{\lambda }(x) = \Psi _{\lambda }(x) \rho _*^{1/2}(x)$.  The probability density  $\rho(x,t)$, that {\it  can be expanded} into  $\phi _{\lambda }(x)$,
  will relax   exponentially  with rates determined by gaps in the energy spectrum of $\hat{H}$.
   This  is what we call the  {\it spectral relaxation   pattern}, c.f. \cite{jph,pavl}.\\

{\it Remark:} As a side comment  let us add that, while  reintroducing  the (purely numerical, dimensionless) diffusion coefficient  $D \neq 1$,  i.e. executing  $\Delta  \to D \Delta $  in Eq. (1),  we  need to pass to  $\rho _* \sim \exp (-U/D)$, and  $b= D \nabla \ln \rho _*$, which gives  Eq. (2)
the form ${\cal{V}}= D [\Delta \rho _*^{1/2}]/\rho _*^{1/2} =    (1/2)(b^2/2D + \nabla b)$.  See e.g. \cite{jph}, subsection A.3, for  a  comparative discussion of the harmonic attraction,   with $D=1/2$ and $D=1$.\\

For concreteness, let us invoke the commonly employed  in the literature higher degree monomial   potentials  $U(x)= x^m/m,  x^m, mx^m$. In   a compact notation we have:
\be
U(x)= {\frac{\kappa_m}m} x^m  \Longrightarrow   {\cal{V}}(x)= {\frac{\kappa _m}2} x^{m-2} \left[ {\frac{\kappa _m}2} x^m - (m-1)\right].
\ee
The  choice of $\kappa _m =1$, $m$ or $m^2$ reproduces  the above listed  functional forms of $U(x)$, in conjunction with  the corresponding   potentials ${\cal{V}}(x)$.
This in turn yields another  parametrization of the potential:
\be
 {\cal{V}}(x) = ax^{2m-2} - bx^{m-2},
\ee
where  $a= \kappa _m^2/4$ and   $b= \kappa_m  (m-1)/2$.

Accordingly, ${\cal{V}}(x)$  has a definite higher order ($m>2$)  double-well  structure,     with two   degenerate  symmetric  minima  at which the potential takes negative values. A local maximum  of the potential  at $x=0$,   equals zero.

 Let us recall that in  the familiar case of the quartic double-well  $ \alpha  x^4-  \beta x^2$, $\alpha , \beta >0$   (which is not in the family (3)), one  is vitally interested in  the existence of bound states related to negative eigenvalues of  $\hat{H}$, see e.g. \cite{jph,turbiner,turbiner1}.   This is not the case, by construction,  for  our higher order double-well systems, where the zero eigenvalue is the lowest isolated  one in the   nonnegative   spectrum of $\hat{H}$.

 The   ground state function     $\rho_*^{1/2}(x)= (1/\sqrt{Z})  \exp [U(x)/2]$  of $\hat{H}$, Eqs. (1)-(3),  is   unimodal    with  a  maximum at an   unstable equilibrium point  of the potential ${\cal{V}}(x)$   profile.   Thus, in the present case,  the  preferred location of the  diffusing (alternatively - quantum)  particle is  to reside   in the vicinity of the  unstable extremum  of ${\cal{V}}(x)$. That is  contrary to  physical  intuitions  underlying the casual understanding  of  tunnelling phenomena in a quartic   double-well  quantum system, c.f. Chapter 4.5 in Ref. \cite{bas}, see also \cite{baner}.

It is worthwhile to mention that the  instability  of the local  maximum of the potential profile,  has been  identified   as a   source of  computational problems  in the  study of  spectral properties of   the quartic double-well  system, for energies close to to the local maximum.  These  have been  partially    overcome  in Ref. \cite{turbiner1} by invoking non-perturbative methods. On the other hand, the quartic  double-well  system   has received an    ample  coverage   in the literature, mostly in connection with   the  tunneling-induced spectral splitting  of    eigenvalues,  located   below the local maximum value and close to this of the    local minimum  (stable extremum of the potential at the bottom of each well).

For  higher-order   double-wells of the form (3), (4), with a local maximum at zero, there are  no negative eigenvalues  of $\hat{H}$  in existence, \cite{jph}, while the existence of the positive part of the spectrum may be considered to be granted in the superharmonic regime.
We note that  Eqs. (1)-(4)  provide  explicit examples  of spectral problems,  for which  neither  a standard semiclassical (WKB) analysis, nor   instanton calculus   (both routinely  invoked in the quartic  double-well case)  can be applied straightforwardly to evaluate    non-zero  eigenvalues of $\hat{H}$, or lowest energy gaps,   \cite{turbiner,turbiner1}.

We are motivated by the   strategy  of Refs. \cite{streit}-\cite{faris}, of reconstructing the (diffusive) dynamics from the eigenstate (and in particular, from the ground state function) of a given self-adjoint Hamiltonian (energy operator). However, in the present paper we follow the  reverse  logic, with  the stochastic process  given a priori  and   the energy operator  and its spectral properties remaining to be deduced. See e.g. an introductory discussion in Ref. \cite{turbiner}. For the uses of eigenfunction expansions in the construction  of transition probability densities of the  pertinent diffusion process, see \cite{risken,pavl,jph}.

Our departure point is the  confining  Smoluchowski process with  a predefined  drift function,  specified  type of noise  (Brownian motion, e.g. the Wiener process) and  an asymptotic   stationary  probability density  function  (pdf) in existence.   In  conformity with Eqs. (1) and  (2), the   {\it zero energy}   eigenfunction  of $\hat{H}$ is  directly inferred  (take a square root)  from the  Boltzmann  equilibrium pdf  $\rho _*(x)$   of the Smoluchowski process.   The    potential ${\cal{V}}$, Eq. (2),  derives from  the  knowledge of  $\rho _*^{1/2}(x)$ alone.

   The problem is that,  for   monomial  attracting potentials  $U(x)\sim x^m$  (with drifts of the form $b(x)\sim - x^{m-1}$),   we   do not know strictly positive  eigenvalues in the discrete spectrum  of the inferred  Schr\"{o}dinger-type  operator $\hat{H}$, nor the related eigenfunctions (c.f.  for comparison,  a  discussion of  sextic and decatic potentials  in Refs. \cite{turbiner,brandon,maiz,okopinska}).   No   reliable  numerical procedures have been developed  to this end, and even for   moderately  large  values of $m$ (say  $m \geq 100$,  known methods (including the Mathematica 12 routines) generally fail.

 To establish  the relaxation properties  (like the time rate of an approach to equilibrium) of the  Smoluchowski process,   we  definitely  need  to have in hands  several  exact or approximate  eigen-data  (basically  energy gaps), at the bottom  of the nonnegative  spectrum  of $\hat{H}$.    This is the essence of the eigenfunction expansion method, \cite{risken,pavl}, while employed  in the study of  asymptotic properties  (e.g.  the spectral relaxation) of  diffusion processes.

The solvability (in the least the partial one) of  involved  Schr\"{o}dinger-type spectral problem appears to mandatory to justify the hypothesis that actually the pertinent diffusion process equilibrates according to the spectral relaxation scenario, see also \cite{dubkov,khar,sokolov} and  earlier research on related  topics \cite{risken}, \cite{kampen}-\cite{liu}.

 In below we shall address this spectral problem, by resorting to approximations  of  the   higher order double-well (3) by  a properly tuned  rectangular double-well,  \cite{blinder,lopez,jelic}.  We have benefited from  Wolfram Mathematica 12    routines of Ref. \cite{blinder}, which  provide   reference  eigendata  (eigenvalues and eigenfunctions shapes of the  energy operator
  $\hat{H}_{well}$),    in  the low part of the  rectangular well  spectrum.   The  steering  parameters of the numerical routine   have   tuning  options, which allow  to  adjust   optimal values of  the interior barrier width and size, once the overall  (sufficiently large, but finite)  height  of the  double-well boundary walls  is selected.

  We have   numerically  tested   circumstances under which the nonnegative  spectrum  actually appears  in the vicinity of the potential  barrier height value.  In  such case, the corresponding  ground state function is predominantly  flat and nearly  constant   (in the least up to the  barrier  boundaries).  This property is     shared by our   higher order double-well systems (conspicuously with the growth of $m$).

   Of special relevance for the approximation procedure   is  that  for $U(x)= m x^m$  (for $U(x)= x^m $ as well),   the local minima of the    inferred potential  ${\cal{V}}(x)$  reside in the interior of $[-1,1] \subset R$  for all $m=2n>2$,  which enables the   fitting  of ${\cal{V}}(x)$  to the rectangular double well  potential  contour, up to an additive "renormalization" of the bottom energy of the  rectangular  well.  The procedure cannot be straightforwardly extended to encompass the case of $U(x)= x^m/m$, $m>2$,  for which the local minima of  ${\cal{V}}(x)$  are exterior to the interval $[-1,1]$, see e.g.  \cite{jph}.

\section{Monomial  potentials  and  the induced  superharmonic  double-wells.}

\subsection{Properties of $U(x)$,   $\rho _*(x)$ and $\nabla \rho _*^{1/2}(x)$   in the  large $m$ regime.}

It is  a folk wisdom, that a sequence of  potentials
$U_m(x) = (x/L)^m/m$, $L>0$, $m$ even, for large values of $m$, can be
used as an approximation of the  infinite well potential  of width $2L$  with {\it reflecting} boundaries located at $x = \pm L$, c.f. \cite{dubkov,khar,dybiec,dybiec0}.
Actually, quite often reproduced  statement reads:  "the reflecting boundary can be implemented by considering the motion in a bounding potential $\lim _{m\to \infty } U_m(x)$",
\cite{khar,dybiec0,dybiec}).  Things are however not that simple  and obvious, see e.g.  \cite{jph} and references  \cite{linetsky,bickel,pilipenko} on the reflected Brownian motion in a bounded domain.

For computational simplicity, we  prefer   to use  the dimensionless  notation $x/L \rightarrow
x$, next  skip the lower index $m$, so  passing  to   $U(x) = x^m/m$.  The ultimate limiting case (e.g.  that of  the  interval with conjectured  {\it reflecting} boundaries) is to be supported on the interval $[-1,1]$.
Potentials of the form $U(x) =\kappa _m  x^m/m $,  Eq. (3), with $\kappa _m= 1, m,  m^2$,       are  employed to the same  end, \cite{khar,dybiec0,dybiec}.

We point out that one needs to observe possibly annoying boundary subtleties. Namely, at
$x = \pm 1$,  we have  the following  limiting values   of $U(\pm 1)$ (point-wise limits, as $m$ grows to $\infty $)
\be
  U(x)= {\frac{\kappa_m}m}  x^m   \Rightarrow  U(\pm 1)= \left \{ {\frac{1}m}, 1, m\right \}   \rightarrow   \{ 0,  1,  \infty \}.
\ee

On   formal grounds, with reference to  the  open  interval $(-1,1)$   and the  complement $R\setminus [-1,1]$ of $[-1,1]$,  we have  the point-wise limit $U(x) \to \infty $, for all $|x|>1$, as $m \to \infty $, while   the interior   limit  equals  zero for all $|x|<1$.
In addition to the emergent  exterior boundary data (instead of the customary local ones for the Neumann Laplacian),  we encounter   significant differences in the boundary ($x = \pm 1$)  properties    of  $m\to \infty$ limit of $U(x)$.

These in turn  have an impact on the limiting properties of   $\rho _*(x)$ (and of the prospective ground state function $\rho _*^{1/2}(x)$ of $\hat{H}$  ). The outcome   appears   not to be quite innocent as  far as the domain of $\hat{H}$ is concerned. Specifically, if  the Neumann Laplacian  $\Delta _{\cal{N}}$  is   to be spectrally approached/approximated  in the $m \to \infty $ limit.

We note that,  as $m \to \infty $,  we arrive at    $\rho _*(x)= 0$ for all $|x|>1$, while $\rho _*(x) = 1/ 2$ for all $|x| <1$.   However, the pointwise limit $m \to \infty $  of $\rho _*(\pm 1)$  reads, respectively (in correspondence with that for $U(x)$, Eq. (5)):
  \be
  {\rho_*(\pm 1)}_{m \to  \infty }  = \left \{\frac 12, \frac{1}{2e},  0 \right \} .
\ee
This implies the  Dirichlet-type boundary data for $\rho _*^{1/2}(x)$ at boundaries $\pm 1$ of the
interval $[-1,1]$.  In case of $\kappa _m = m^2$  we deal with  a "canonical" form  of Dirichlet boundaries, since $\rho _*^{1/2}(x)$  vanishes  at $x =\pm 1$.

The behavior of $U(x)$ and $\rho_*(x)$ with the growth of $m$, we depict in Fig. 1 for the case
of $U(x)= mx^m$.  The exemplary  functional forms of the inferred potential ${\cal{V}}(x)$ are shown in Fig.2. The conspicuous higher order double well structure is clearly seen.

\begin{figure}[h]
\begin{center}
\centering
\includegraphics [width=0.4\columnwidth] {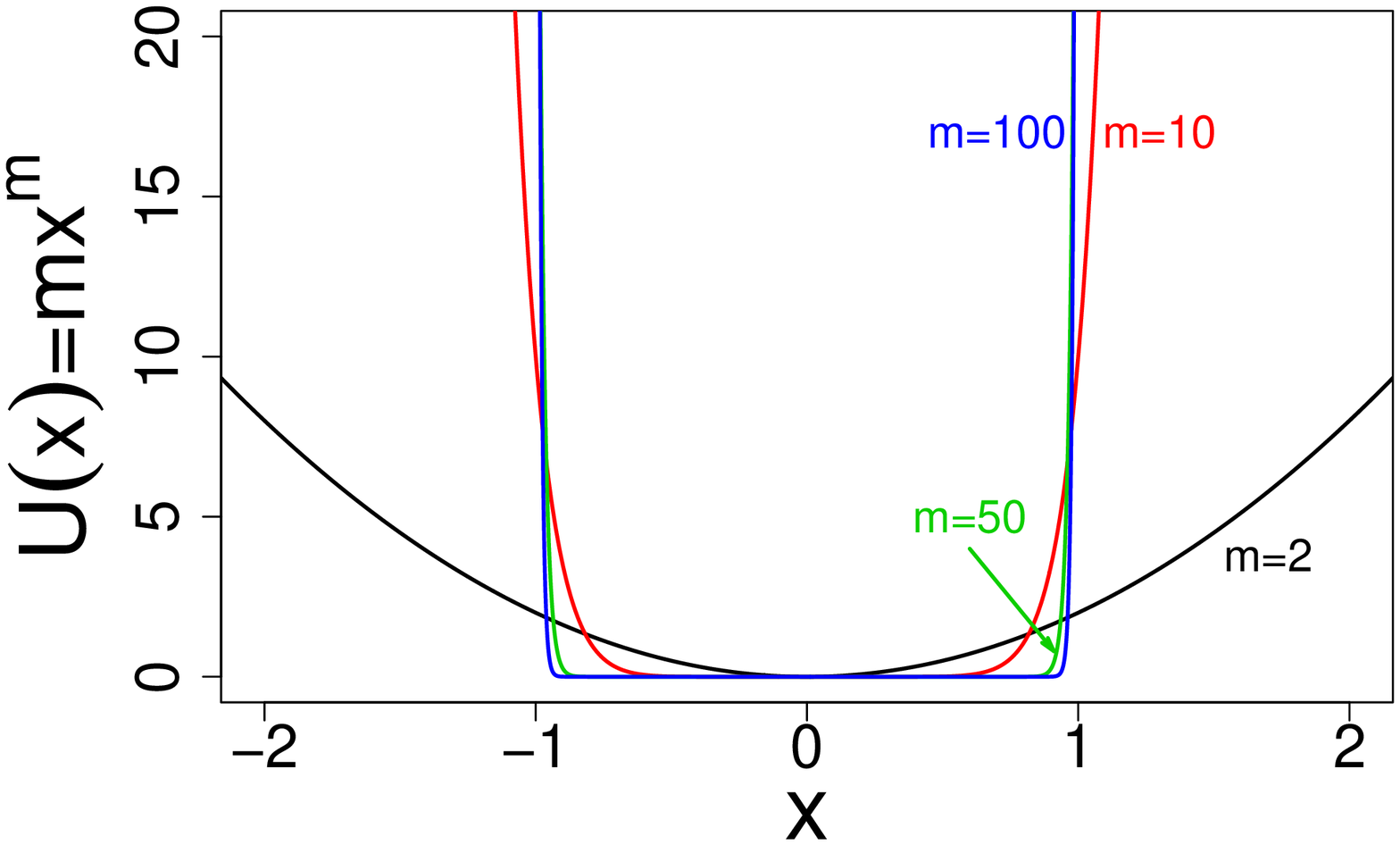}
\includegraphics [width=0.4\columnwidth] {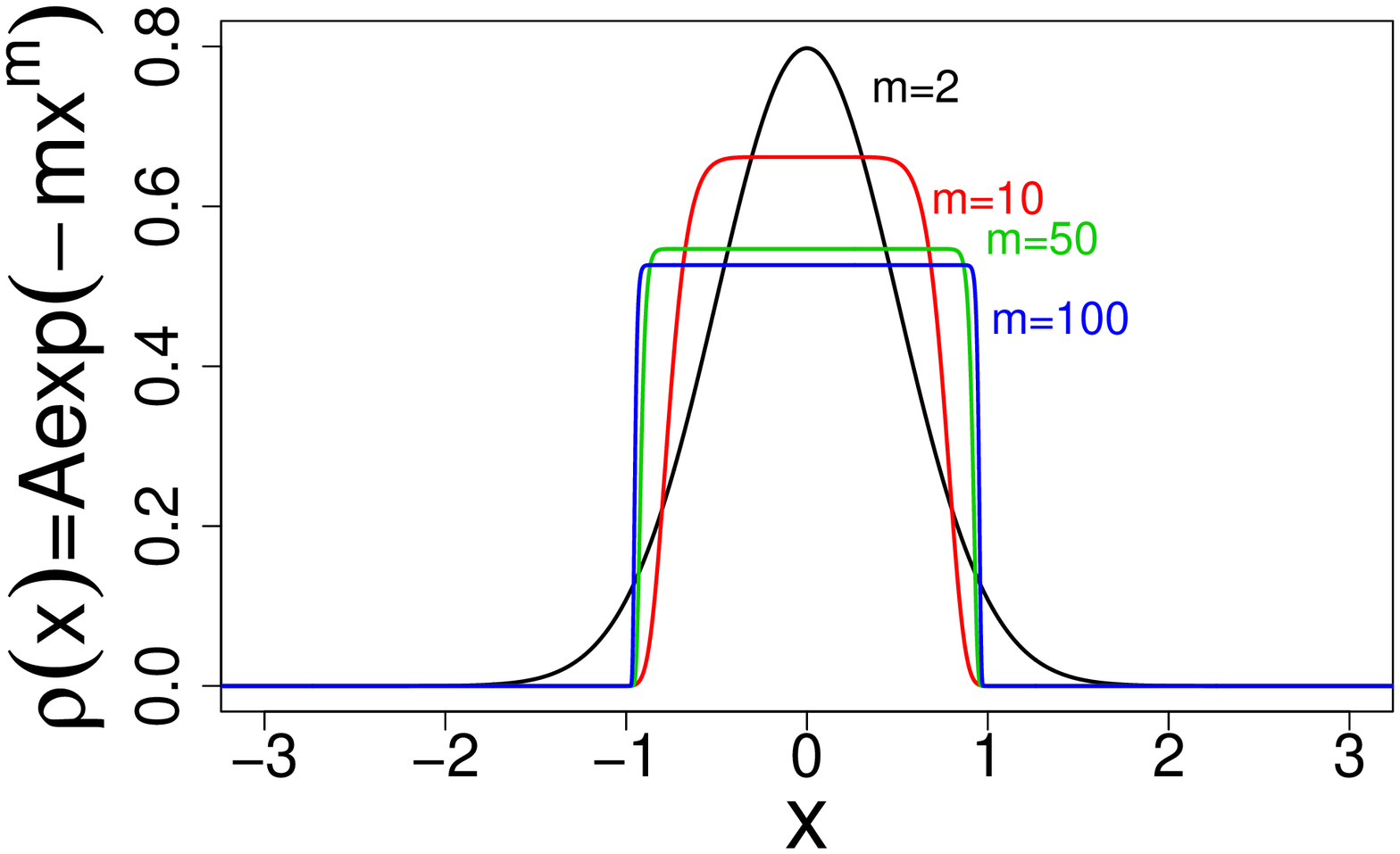}
\caption{Left panel: $U(x)=  mx^m$,  $m=2n>2$.  Right panel: we report  exemplary  stationary pdfs of   the Smoluchowski  diffusion process    $\rho _*(x)= (1/Z) \,  \exp [ - U(x)]$,   $Z  = \int_{-\infty }^{+\infty } \exp[-U(x)] dx$.    We have  $\lim_{m\to \infty } U(x) = \infty $ for all $|x| \geq 1$,  and the limiting value    $0$ for  $|x| < 1$.   Accordingly,   $\rho _*(x)  \to  0$   for  $|x|\geq 1$ and   $ 1/2 $ for all $|x|<1$.  We note symptoms of  the  (graphical) convergence towards the  constant  pdf   $1/2$   in $(-1,1)$, which sets an association with the reflected Brownian motion in  the interval $[-1,1]$, c.f. \cite{jph}. One should carefully observe the troublesome  point-wise limit  (the Dirichlet  boundary) of   $\rho _*(\pm 1)=0 $ as $m  \to  \infty $,  see e.g.  Eq. (6).}
\end{center}
\end{figure}

As long as we prefer to deal with traditional Langevin-type  methods of analysis,  it is  of some   pragmatic interest  to know, how reliable  is  an approximation of the reflected Brownian motion in $[-1,1]$ by means of  the   attractive  Langevin driving (and thence by solutions of Eq. (1)),  with force terms (e.g. drifts)  coming from  extremally anharmonic (steep)  potential wells.

The main obstacle, we encounter here,  is that a "naive" $m=2n \rightarrow \infty $ limit is singular and cannot be
 safely executed on the level  of  diffusions  proper.
 We note that  for any finite $m$, irrespective of how large $m$ actually is,  we deal with a continuous and
 infinitely differentiable higher order double-well  potential and the smooth   Boltzmann- type pdf.  These properties are broken in the (formal) limit $m\to \infty$.

At this point we recall that for  the Neumann Laplacian  $\Delta _{\cal{N}}$   on $[-1,1]$, the boundary condition for all  its  eigenfunctions,  \cite{bickel,carlsaw},   is imposed  locally. In particular, for the ground state function we should have $\Delta  _{\cal{N}} \rho _*^{1/2}(x)=0$  and    $\nabla \rho _*^{1/2}(\pm 1) =0$, see e.g. also  Ref. \cite{carlsaw}, chap. 4.1.
 This  formally  holds true or  a constant function $1/\sqrt{2}$, defined on $[-1,1]$, but can not be achieved through a controlled   limiting procedure:   $m\to \infty $  of  $\nabla \rho _*(\pm 1)$.

Indeed,  we know from Ref. \cite{jph} that for $U(x)= x^m/2$, the inferred square root of the Boltzmann-type pdf  $\rho _*^{1/2}(x)=(1/\sqrt{Z})  \exp(-x^m/2m)$ does not reproduce the Neumann condition   in the $m\to \infty $ limit.  Namely, we have:  $ \lim_{m\to \infty } \nabla  \rho _*^{1/2}(+1) = - 1/2\sqrt{2}$.

Since, in accordance with the notation  (3) we have:
\be
\nabla \rho _*^{1/2}(x)  =   - {\frac{1}2} [\nabla U(x)]    \rho _*^{1/2}(x)
\ee
where $\nabla U(x)= \kappa_m x^{m-1}$,  we can readily complement formulas (5) and (6), by these  referring to   the limiting behavior of  $\nabla \rho _*^{1/2}(\pm 1)$:
\be
{\nabla \rho_* ^{1/2} (\pm 1)}_{m\to \infty} = \left \{\mp {\frac{1}{2\sqrt{2}}}, \mp \infty , \mp \infty \right \}.
\ee
Clearly, the point-wise $m \to \infty $  limit of $\nabla \rho _*^{1/2}(\pm 1)$  has nothing to do with the Neumann boundary condition,  which is {\it   not reproduced}   in  this  limiting   regime. Conseqeuntly,    $\rho _*^{1/2}(x)$  is  {\it  not} in  the domain of the Neumann Laplacian  $\Delta _{\cal{N}}$, in plain contradiction with   popular expectations,  \cite{dubkov} - \cite{dybiec0}.

\begin{figure}[h]
\begin{center}
\centering
\includegraphics[width=0.4\columnwidth]{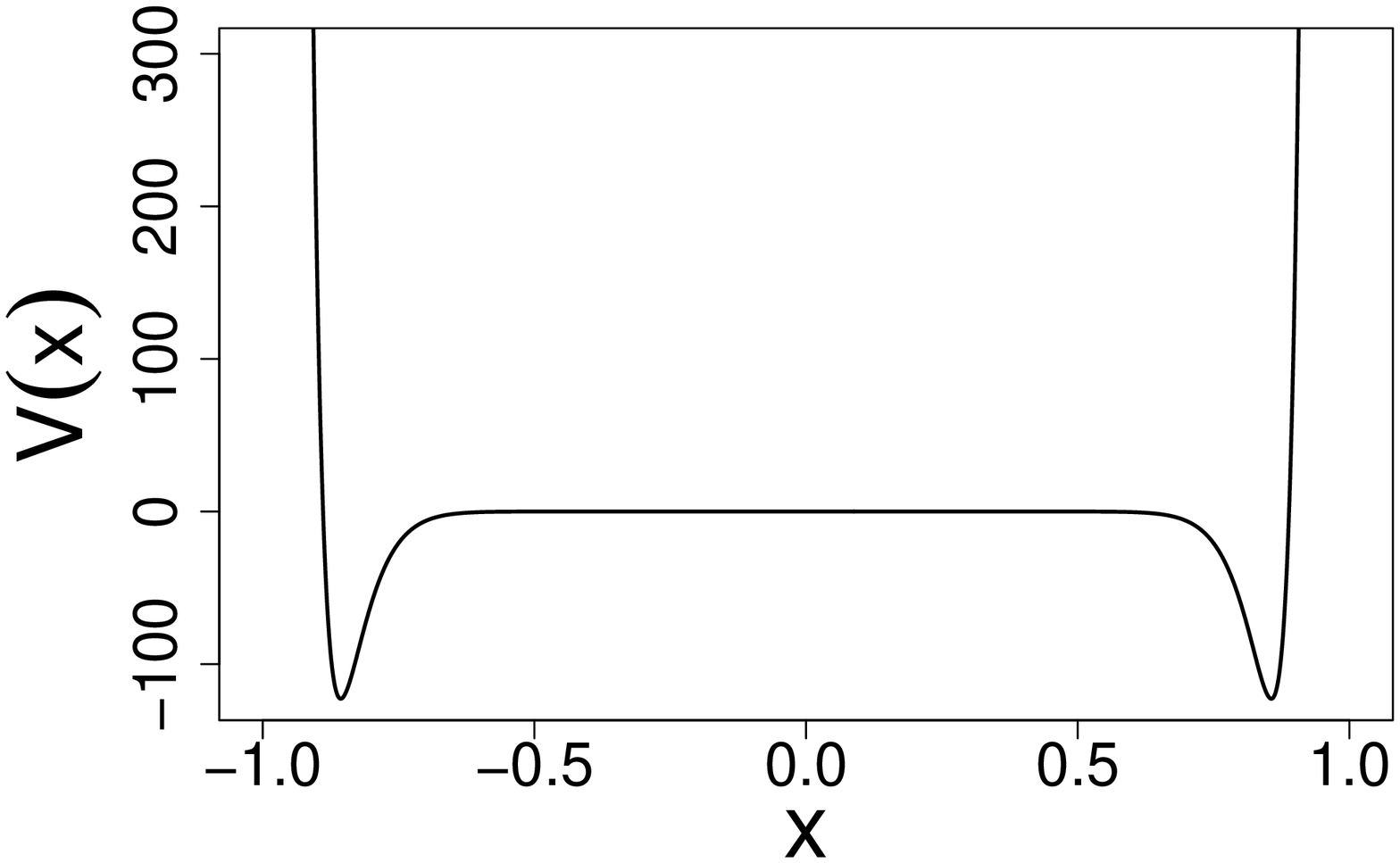}
\includegraphics[width=0.4\columnwidth]{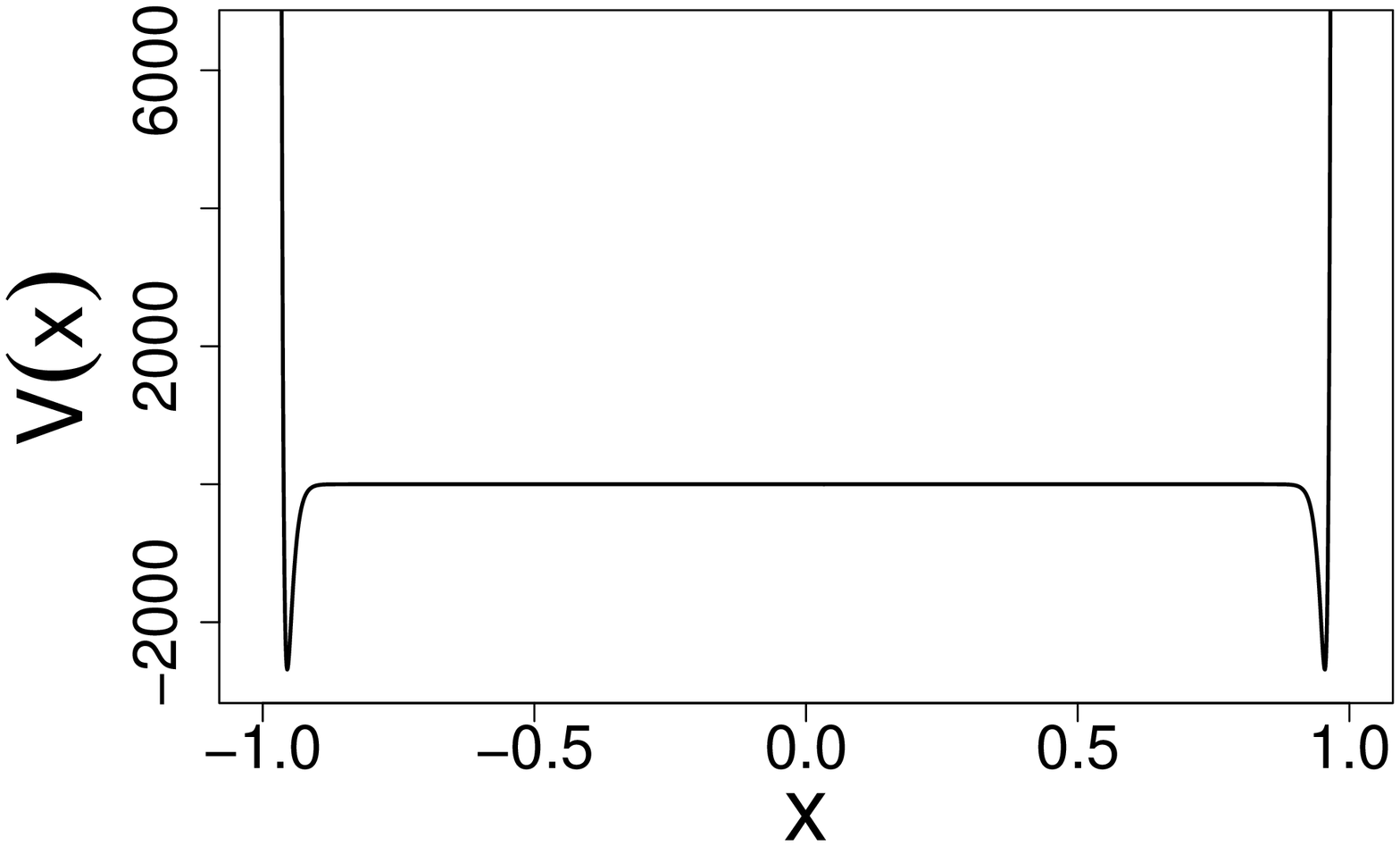}
\caption{ $U(x)= m x^m$: the inferred  potential ${\cal{V}}(x)
= {\frac{m^2}2} x^{m-2} [{\frac{m^2}2} x^m +(1-m)]$  is
depicted  for $m=20$ and $100$.  Note  significant  scale differences along the vertical axis.
 Minima  of  the semigroup potential are located in the interior of $[-1,1]$  for all $m$,  and with the growth of $m$,   approach  $\pm 1$.  In parallel, the depth of  narrowing  local wells  escapes to minus infinity, while the  ${\cal{V}}(x)=0$  "plateau" extends to  $(-1,1)$.   We note that for all $|x| \geq 1$   we have   ${\cal{V}}(x) \rightarrow + \infty $.}
\end{center}
\end{figure}

\subsection {Location of the minima $|x_{min}|$ of   ${\cal{V}}(x)$ and the  large $m$ asymptotics.}

We can readily infer the location of  (negative) minima of the potential ${\cal{V}}(x)$, c.f.  Eq. (3) and   Fig. 2,  where
\be
\mathcal V'(x)=0   \Longrightarrow    |x_{min}|= \left[ {\frac{b}{2a}} {\frac{m-2}{m-1}}\right]^{1/m}  =\left[ {\frac{ m-2}{\kappa _m}} \right]^{1/m}.
\ee
For $\kappa _m=1$ we have $|x_{min}|= (m-2)^{1/m}>1$ for all $m>2$. For $\kappa_m =m$ we obtain  $|x_{min}| = [(m-2)/m]^{1/m}<1$, and  likewise  for $\kappa_m=m^2$, when  $|x_{min}|= [(m-2)/m^2]^{1/m}<1$.

 We  point out that $m^{1/m}>1$ and   $\lim _{m\rightarrow \infty } m^{1/m} =1$, c.f. \cite{kuczma,jph}.  Accordingly, in the large m limit, the minimum locations  approach  the  interval $[-1,1]$ endpoints  $\pm 1$, respectively   from the  interval  exterior  $R\setminus [-1,1]$ for $\kappa_m=1$,   or interior  of  $[-1,1]\subset R$ if otherwise. This   behavior is  (continuously)  depicted   in  Fig.3.

\begin{figure}[h]
\begin{center}
\centering
\includegraphics[width=0.5\columnwidth]{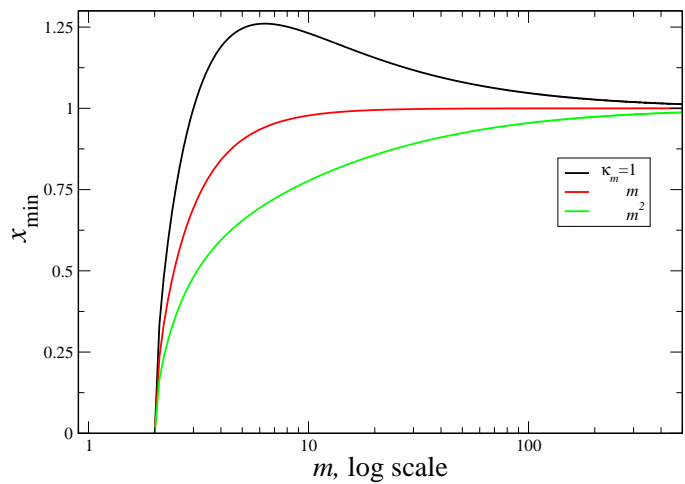}
\caption{Dependence of $|x_{min}|$ on  $m$  (here visualised continuously),   for different choices of  $\kappa _m=  \{1, m, m^2\}$.}
\end{center}
\end{figure}

Since we  are interested in the large m regime it is useful to rewrite the formula  (9) as follows:

\be
|x_{min}| =  \exp \left \{ {\frac{1}m}  \left[ \ln {\frac{m}{\kappa_m }}  + \ln (1 - {\frac{2}m}) \right] \right \}
\sim  1+\frac{\ln (m/\kappa _m)}{m}-\frac{2}{m^2}+\frac{\ln^2(m/\kappa _m)}{2m^2},
\ee
where $\ln (m/\kappa _m)= \{\ln m, 0, - \ln m \}$,  and we have employed the series expansion $
\ln (1-x)=  -  \sum_{n=1}^{\infty }  x^n/n$,   valid  in the  range $-1 \leq x <1$, and  here considered for $x= 2/m$.   In the   large $m$ approximate formula (10), expansion terms up to the $m^{-2}$ order have been kept.  We recall that $\kappa_m=\{1, m, m^2\}$.

We immediately realize that in the regime of large $m$, for  $\kappa _m=1$ the dominant contribution  to $|x_{min}|$, Eq. (10),  comes from $1+ {\frac{\ln m}{m}}$,   for $\kappa_m=m$  from $1- 1/m^2$, and for $\kappa_m= m^2$ from  $1- {\frac{\ln m}{m}} $.

Let us  denote
\be
\triangle   = \triangle (m) = |1- |x_{min}||
\ee
the  distance   of $|x_{\min}|$ from the nearby boundary point   $\pm 1$. In passing we note that $\triangle  \sim \{ +   {\frac{\ln m}{m}}, 1/m^2,  {\frac{\ln m}{m}} \} $ for $\kappa_m= \{1, m, m^2\}$ respectively.

 In Fig. 4  we   visualize  the $m$-dependence of the {\it signed}  deviation   $  \delta  =  1- x_{min}$  of $+1$  from   the nearby  $x_{min} >0$.    For large $m$, we have  $\delta   \sim \{ - {\frac{\ln m}{m}}, 1/m^2, {\frac{\ln m}{m}} \}$.   For $x_{min}>1$ the  signed deviation is negative, and positive for $x_{min}<1$.  Note that $\triangle= |\delta |$.

\begin{figure}[h]
\begin{center}
\centering
\includegraphics[width=0.5\columnwidth]{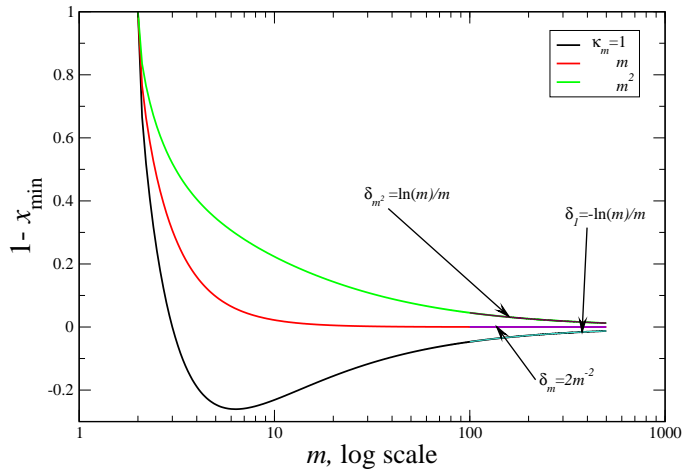}
\caption{Dependence of $\delta =1- x_{min}$  on $m$, with  $x_{min}>0$,    while  visualised continuously  for different choices of  $\kappa _m=  \{1, m, m^2\}$. }
\end{center}
\end{figure}

\subsection{Variability of ${\cal{V}}(x)$  in the vicinity of   $x=\pm 1$.}

By turning back to Eq. (3), plugging there the minimum location value $|x_{min}|$,  Eq. (9), we arrive at the following expression for  the depth of local wells of the potential (3):
\be
{\cal{V}}(|x_{min}|) = - {\frac{m\kappa_m}4} |x_{min}|^{m-2} = - {\frac{m(m-2)}4} |x_{min}|^{-2}.
\ee

 \begin{figure}[h]
\begin{center}
\centering
\includegraphics[width=0.5\columnwidth]{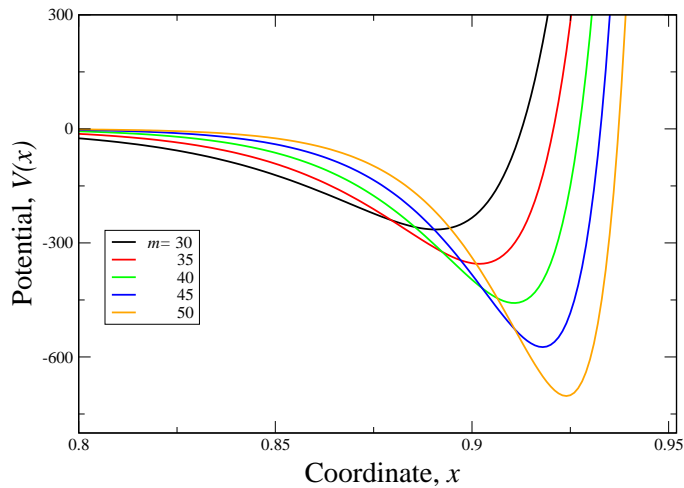}
\caption{A complement to Fig. 2, with  ${\cal{V}}(x)$ inferred from $U(x)= m x^m$.  Sequential image of local well minima in the vicinity of  $x=1$, for moderate values of     $ m=  30, 35, 40, 45,, 50$.   A  flat part of the potential  curve  ${\cal{V}}(x)  \simeq     0$ extends (collectively)  between  rough values $-0.8$ and $+0.8$. Note that for $m=50$, we have  $|x_{min}|\simeq  0,92$,  $2\triangle \simeq 0,16$ and the  flat part of the  potential  curve  is  roughly limited to   $[-0,84, +0,84]$.}
\end{center}
\end{figure}

For sufficiently large values of $m$  (basically above $m=100$),   we can pass to  rough approximations
\be
 {\cal{V}}(|x_{min}|) \sim - {\frac{m(m-2)}4} \sim   - {\frac{m^2}4}.
\ee
  (We note that in this  approximation regime,  the well depth becomes independent of the choice of $\kappa _m$.) This rough estimate of ${\cal{V}}(|x_{min}|)$, comes from  presuming   that  $|x_{min}|^{-2}  \to 1$.

  It is instructive to  have more detailed insight into  the pertinent large $m$ asymptotics.
We proceed  by  repeating  basic steps in the derivation of   Eq. (10):
\be
|x_{min}|^{-2}=  \exp \left \{- {\frac{2}m}  \left[ \ln {\frac{m}{\kappa_m }}  + \ln (1 - {\frac{2}m}) \right] \right \}
\sim    1- \frac{2\ln (m/\kappa _m)}{m}+\frac{4}{m^2}+\frac{2\ln^2(m/\kappa _m)}{2m^2}.
\ee
For large $m$, the dominant contributions read: for $\kappa = 1$ we have $1- 2{\frac{\ln m}{m}}$,   for $\kappa_m=m$  we get   $1+ 2/m^2$, and for $\kappa_m= m^2$  we have   $1+ 2{\frac{\ln m}{m}}$.
This outcome  lends support to our  approximation (13)   of the   well depth.

We recall that for sufficiently large values of  $m$,   local  minimum  locations    $x_{min}$ are close to $\pm 1$, and in the interval of  the  size $2\vartriangle $ in the vicinity of $\pm 1$, we encounter rapid  (albeit smooth, e.g.  continuous and continuously differentiable) variations of  ${\cal{V}}(x)$.
 Considering $m$ to be large,  we exemplify  this behavior  for $\kappa_m=m^2$, within  the  interval $1 - 2\triangle <  x_{min} <1$   of length $2\triangle $:
  \be
  {\cal{V}}(x_{min}- \triangle ) \simeq 0  \Rightarrow {\cal{V}}(x_{min}) \simeq  - {\frac{m(m-2)}4}
  \Rightarrow   {\cal{V}}(x_{min}+ \triangle )= {\cal{V}}(1)=  {\frac{m^2}2}\left[ {\frac{m^2}2} - (m-1)\right]    \simeq   {\frac{m^4 - 2m^3}4},
\ee
We have  thus a  "wild"   variation  of ${\cal{V}}(x)$,   ranging from  nearly  $0$,  through  (roughly)  $- (m^2  - 2m)/4$, to  (roughly as well)  $+(m^4 - 2m^3)/4$,  in the interval of length $2\triangle \sim  2/m^2$.

\section{Rectangular double-well approximation of the  superharmonic  double well potential ${\cal{V}}(x)$.}

\subsection{The fitting procedure.}

For further discussion we restrict consideration to the choice of $U(x)= m x^m$, with all ensuing formal consequences.  Essentially, we need $|x_{min}| <1 $ for all $m$, an  the  point-wise  limit
of $\lim_{m\to \infty } U(\pm 1) = \infty $.   Since  $\rho_*^{1/2}(\pm 1)_{m\to \infty } =0$, we are tempted to explore  an approximation of our higher degree double-well potential function, by means of a  sequence  of    rectangular  double well systems, with  adjustable  (internal)  barrier  heights and widths, \cite{bas}.

 We anticipate  the existence of an affinity     between  ${\cal{V}}(x)$,  as $m$ grows to $\infty $,   and  a  properly tuned    rectangular double-well potential,  cf. Refs. \cite{bas, blinder, lopez,jelic}.   That  is supported  by an  experimentation  with the  dedicated Mathematica 12 routine,  \cite{blinder},  created to address the rectangular  double-well spectral problem.  Fine tuning options concerning the  overall  depth  (large)  of the well and the middle barrier size parameters  (width and height) allow  for  a controlled manipulation. Its explicit outcome were lowest eigenvalues and eigenfunctions  (up to eight) of the  corresponding energy
  operator $\hat{H}_{well}$, see e.g. \cite{blinder}.

 Numerical tests confirmed  that the ground state eigenvalue, which is  equal (or in the least nearly equal)  to the height of the barrier,   is in existence   if the  proper  width/height balance is set. The corresponding ground state eigenfunction is   "flat" (practically constant) in the area of the barrier plateau  (local maximum area).  This  sets a background for a subsequent discussion.

We depart from the "canonical"  qualitative picture of the ammonia molecule, as visualized in terms
of the rectangular double well, \cite{bas}, chap. 4.5. We adopt the original notation  of Ref. \cite{bas} to the graphical description  given  below,  in Figs 6  and 7, see also \cite{blinder}. One must  keep in mind  different (D=1 versus D=1/2) scalings of the Laplacian in the employed   versions of the   rectangular well energy operator.

From the start, we implement the energy scale "renormalization" of the rectangular double-well  energy operator.  The original non-negative potential \cite{bas,blinder,lopez,jelic}   is shifted down along the energy axis, by the value of its   local maximum  (barrier height),  from the original minimum value  $0$ (well bottom)  of the  rectangular potential.  This in turn    enables   and effective  fitting of  the higher order double-well  potential to the rectangular double-well potential contour.

\begin{figure}[h]
\begin{center}
\centering
\includegraphics[width=0.5\columnwidth]{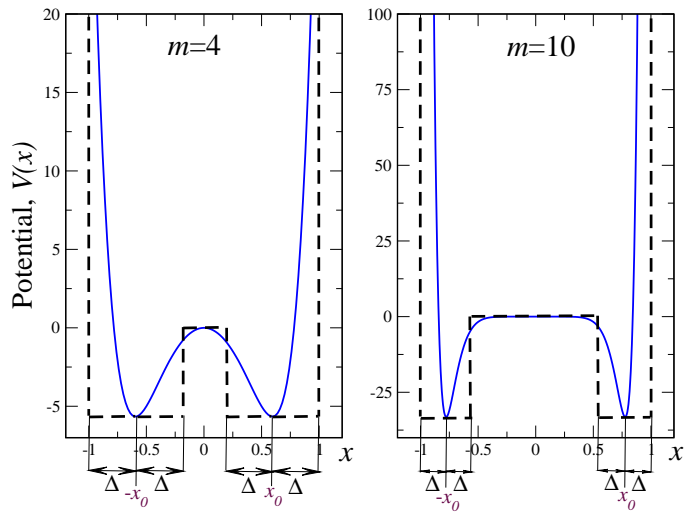}
\caption{Exemplary fitting of $m=4$ and $m=10$  potentials ${\cal{V}}(x)$  to  suitable rectangular double wells. We point out the   overall  shift of the original rectangular  double-well  along the energy axis. This  assigns the value $0$ to the local maximum (top of the  interior  barrier)  and makes the  height of the barrier to quantify the well depth.   Here, the interval in use is $[-1,1]$,  $\triangle $ stands for a distance between the nearby endpoint $\pm 1 $ and  the  location $\pm x_0  = x_{min}$  of the  local minimum of ${\cal{V}}(x)$. Here $2 - 2\triangle $  is a provisional measure of the "plateau"  width,  of the  (fitted) rectangular well barrier.}
\end{center}
\end{figure}

\begin{figure}[h]
\begin{center}
\centering
\includegraphics[width=0.55\columnwidth]{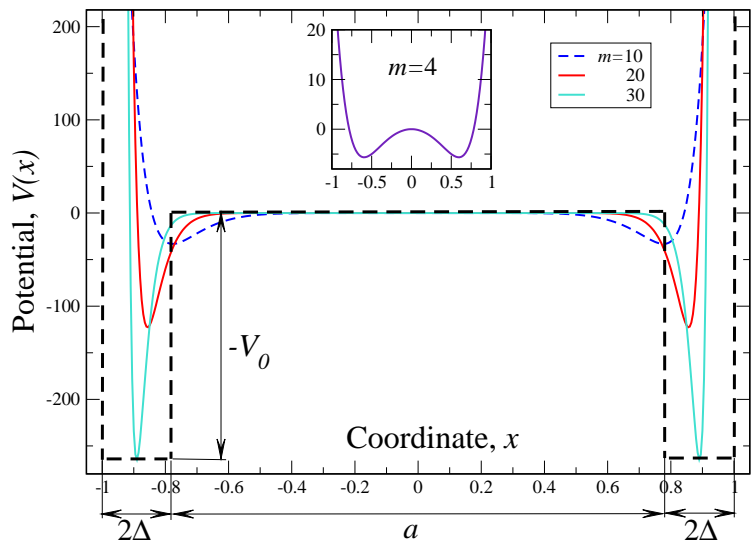}
\caption{A comparative display  of $m=10, 20, 30 $  potentials ${\cal{V}}(x)$, where $x_0$,   $2\Delta$  and the height $V_0$ of the barrier  ($- V_0$ refers to  the  depth of the well),  actually provide a fit for  $m=30$.  The  $m=4$ potential contour   is depicted in the inset.}
\end{center}
\end{figure}

The fitting procedure, graphically outlined in Figs. 6 and 7,  looks  promising  but is somewhat deceptive.  To justify its usefulness, we must first check under what circumstances the rectangular well spectral problem does admit the bottom (ground  state) eigenvalue zero.  In contrast to the higher order double-well $\hat{H}$, Eqs. (1) to (3), where the zero energy ground state is  introduced as a matter of  principle,  its existence is obviously {\it not } the   generic  property  in  the   rectangular double-well setting. Accordingly, the proposed approximation methodology might seem bound to fail.

Fortunately, we can demonstrate that   potentially disparate two-well   settings  (higher order versus rectangular one), actually  coalesce if we look   comparatively  at the higher $m$  data (say $m\geq 80$) and set them in correspondence with these belonging to the rectangular well system. To this end, let us employ the rectangular double-well lore of Ref. \cite{blinder}.  Temporarily, the notation will at some points differ from this adopted by us in Section  III.A.

Following Ref. \cite{blinder}, we  consider  the potential $V_{rect}(x)= V(x)= \infty $ for $x<0$ and $x> \pi $, while we assign a constant positive value $V(x)=V_0$ for $(\pi -a)/2\leq x \leq (\pi +a)/2$,
where $0<a<\pi $,  and demand $V(x)=0$ elswhere.  This defines a rectangular double well profile immersed in the infinite well. The  double well is set on the interval $[0,\pi ]$, its bottom is located at the energy value zero, while the barrier with height $V_0$ has the width $a$, and separates two symmetrical wells  extending  over  the intervals $[0, (\pi -a)/2]$  and  $[(\pi +a)/2, \pi ]$.

\subsection{The eigenvalue zero in  the rectangular double-well  setting.}

As far as the ground state function and  the bottom eigenvalue of   $\hat{H}_{rect} = -{\frac{1}2} \Delta + V(x)$  is concerned,  we have in hands two steering parameters $a$ and $V_0$, which can be fine-tuned.  In passing, we notice the presence of the $1/2$ factor preceding the Laplacian, and  recall  that the interval of interest  is $[0, \pi ]$ instead of $[-1,1]$.   This needs to be accounted for, when we shall pass to  the spectral comparisons between  $\hat{H}_{rect}$ and $\hat{H}$ of   Eqs. (1)-(3).

In view of the standard infinite well enclosure,  all   (piece-wise connected)  eigenfunctions are presumed to obey the  {\it  Dirichlet boundary data}: $\psi (0) = 0 = \psi (\pi )$.
We  are interested in the ground state function, hence our  focus  is  on  even eigensolutions  $\psi(x)= \psi(\pi - x)$  of $\hat{H}_{rect} \psi  = E \psi $.

In the two local  well areas we have  $V_0(x)=0$ , hence respective even solutions have the  self-defining form, \cite{blinder,bas}:
\be
\psi _L(x)= \alpha  \sin(kx), \, \,  \,   0\leq x\leq (\pi -a)/2
\ee
and
\be
\psi _R(x)= \alpha \sin [k(x- \pi )],  \, \, \,  (\pi -a)/2 \leq  x\leq  \pi
\ee
where  subscripts  $L$ and  $R$  refer to  the   left or right well, respectively.
Within the wells we have:
\be
- {\frac{1}2} {\frac{\Delta \psi (x)}{\psi (x)}}= {\frac{k^2}2} = E.
\ee

On the other hand, within the barrier region,  the proper form of the  even eigenfunction is:
\be
\psi _{barrier}(x)= \beta \cosh [{\cal{K}}(x - \pi /2)], \, \, \, (\pi -a)/2 \leq x\leq (\pi +a)/2.
\ee

Since the  total energy is preserved  throughout the well and
equal $E$, Eq. (18),  along the barrier we have:
\be
\left[- {\frac{1}2} \Delta  + V(x) \right] \psi (x)   = \left[ { \frac{{\cal{K}}^2}2} + V_0\right] \psi (x)  = E \psi (x).
\ee
Accordingly, for $E \geq V_0 $, we have
\be
{\cal{K}}= \sqrt{2V_0 - k^2}=\sqrt{2(V_0 - E)}.
\ee

The connection formulas at  the barrier  boundaries,  may be conveniently expressed as continuity conditions for logartithmic derivatives. For example at $x= (\pi -a)/2$ we  require:
\be
\nabla \ln  \psi _{well}([\pi \mp a]/2) = \nabla \ln  \psi_{barrier}([\pi \mp a]/2),
\ee
which results in the transcendental equations (for even states):
\be
k \cot [k (\pi -a)/2]  =   -  \sqrt{2V_0 - k^2} \tanh [(a/2)\sqrt{2V_0 - k^2}].
\ee

We point  out, that  the regime of $V_0 \leq E$ may be achieved  a formal substitution ${\cal{K}} \rightarrow   i {\cal{K}}$, where $i$ is an imaginary unit.  This would transform $\cosh({\cal{K}}(x- \pi /2)$ into $\cos ({\cal{K}}(x- \pi /2)$, in parallel  with  the replacement of  $V_0 -E\geq 0$  by $E- V_0\geq  0$  in Eq. (21).

 The transition point  between  two spectral regimes  $E \leq V_0$ and $V_0 \leq E$ follows from the demand:
\be
2V_0- k^2= 0  \Longrightarrow  k \cot [k (\pi -a)/2] =0
\ee
which implies  $k= \sqrt{2V_0}$ and $E=  k^2/2 = V_0$.

The condition (24) sets a relationship between $a$ and $V_0$, showing  for which pairs ${a,V_0}$, the spectral (non-negative) ground state eigenvalue $E= V_0$ is admissible.
We have:
\be
a=a(V_0)= \pi \left( 1- {\frac{1}{\sqrt{2V_0}}} \right)
\ee
or, equivalently:
\be
V_0= V_0(a) = {\frac{\pi ^2}{2(\pi - a)^2}}
\ee
which we depict in Fig. 8.

\begin{figure}[h]
\begin{center}
\centering
\includegraphics[width=0.6\columnwidth]{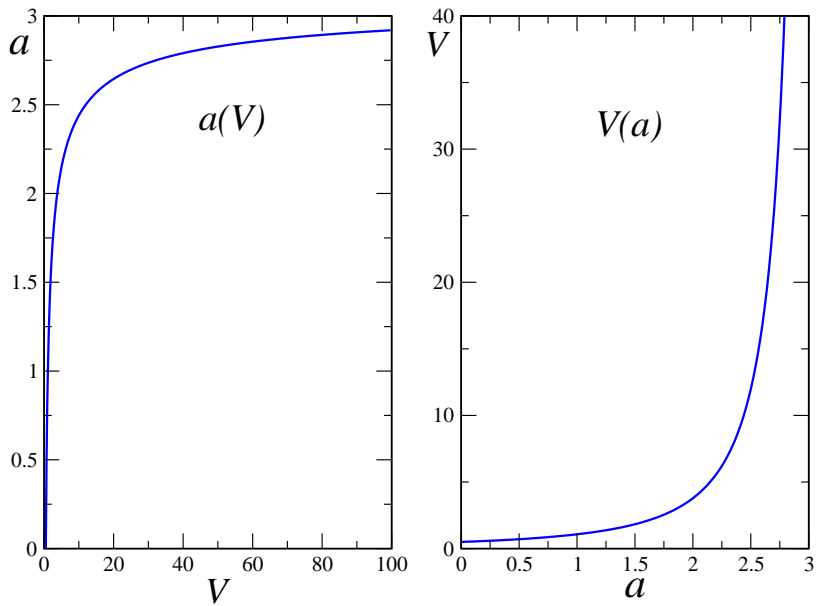}
\caption{In the left panel we depict a dependence of $a$ upon $V_0$, Eq. (25), while meeting the condition (24). In the right panel, we  depict the dependence of $V_0$ upon $a$, according to Eq. (26).}
\end{center}
\end{figure}

We note that the condition (22)  requires ${\cal{K}}=0$, and in agreement with Eq. (19) identifies
$\psi _{barrier}(x)= \beta $  to be  constant  along the barrier "plateau".  Given $a$, we have  in hands  the corresponding barrier height  $V_0=V_0(a)$, Eq. (26). Since $k= \sqrt{2V_0}$, we have also in hands an  explicit functional form for the  left and right  well  eigenfunction  "tails" $\psi _L(x)$ and $\psi _R(x)$, c.f.  Eqs. (16), (17).  We point out the validity of the Dirichlet boundary conditions at $0$ and $\pi $. Moreover, the continuity   conditions (22) for logarithmic  derivatives at the barrier endpoints, do  not need nor necessarily imply the Neumann condition (e.g. the vanishing of derivatives at these points).

Coming back to the  comparative  (higher order well versus rectangular well) "eigenvalue zero" issue, let us notice that plugging ${\cal{K}}=0$ in Eq. (20), we need to "renormalize" the energy scale by subtracting $V_0$:
\be
\left\{- {\frac{1}2} \Delta  + [V(x) - V_0] \right \} \psi (x)   = 0.
\ee
to pass to the "eigen-energy zero" regime of the rectangular well problem.   This is properly reflected in Figs. 6 to 8.  We note, that to maintain the    link with the potential (3)  for all $m$, we   need   to allow $V_0$ to escape to  $\infty $, with  $m  \to \infty $. This  may be considered as a motivaton for invoking the phrase  "additive energy renormalization", in connection with the $V_0$ - subtraction in Eq. (27).\\

{\it Remark:} Let us mention that the  zero energy association with the unstable equilibrium  of the potential profile,  has been discussed for standard quartic double well.  A transition value of the well steering  parameter has been identified, \cite{turbiner,turbiner1,jph},  as a  sharp  divide  point  between two spectral regimes: nonnegative and that comprising a finite number of negative eigenvalues  (near the local minima standard WKB methods give reliable spectral outcomes for the "normal"   double-well system,  \cite{blinder,lopez,jelic}).

\section{Discussion of spectral affinites: Superharmonic double well versus  rectangular double well.}

\subsection{Notational adjustments.}

Since some of the defining   parameters in the rectangular double-well and in the higher order  (superharmonic) double-well  differ,    we need to  analyze  means of the removal of this obstacle  in  our  subsequent analysis.

First we shall   comparatively  address  the width/height  balance of the barrier   in the rectangular case,  with its analog  (approximate width and the elevation of the "plateau" above the potential minima, c.f. Fig. 7)  in the higher order case.  To this end, we need to resolve the $[0,\pi ]$ of  Fig. 8,   versus $[-1,1]$ of  Figs. 6 and 7,    interval  size  discrepancy.

We note that it is $U(x)=(x/L)^m$,   which in the large $m$ regime  gives rise to the well with the support on  the interval $[-L,L]$  of length $2L$. Setting $L=1$ we recover $[-1,1]$, while $L=\pi /2$ gives rise to $[-\pi /2, \pi /2]$.

These support intervals are  examples of  centered  boxes with the center location $x_c=0$.  An arbitrarily relocated  (shifted)  box  of length  $2L$, if centered around  any $x_c \in R$ has a  support $[x_c-L,x_c +L]$.  Hence, choosing $x_c= L$, we pass to the supporting interval $[0, 2L]$, which upon the adjustment $L=1 \rightarrow L = \pi /2$ leads to the required  $[0,\pi ]$.

\begin{figure}[h]
\begin{center}
\centering
\includegraphics[width=0.5\columnwidth]{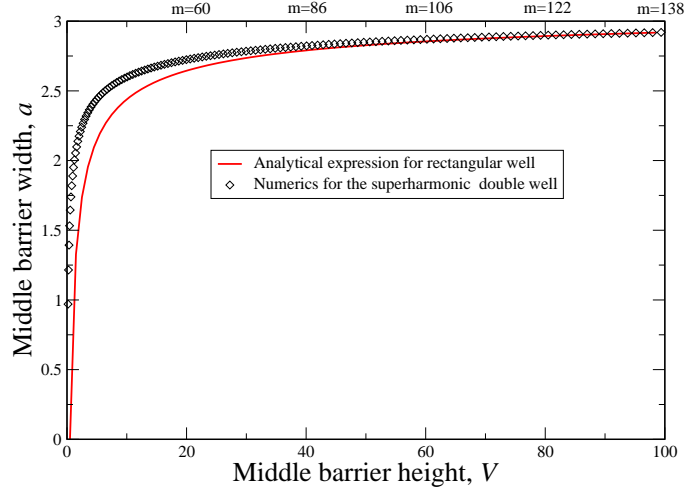}
\caption{After adjustments of  the interval location and length ($[-1,1] \rightarrow [0, \pi]$) and   steering parameters $V_0$,  $a$ (see the text)  we have identified when the eigenvalue zero regime of the rectangular double-well actually coalesces with that of the higher order (superharmonic)  double-well.  A fapp (for all practical purposes) coalescence   can be  (with quite reasonable accuracy)  accepted  for  $m \geq  84$.}
\end{center}
\end{figure}

The  rectangular well width-height/depth, $a$-$V_0$ balance, as depicted in  Figs. 7 and 8, is to be compared  with the   corresponding  data of  the superharmonic   double-well system (1)-(3). To this end,  we must recompute the potential (3) minima (their bottom is  set at $-V_0$  in Figs. 6 to 8), identify the location of $x_0$, and next  evaluate  $4 \triangle $, to  get the width identifier  $a= \pi - 4 \triangle $. The computation must be accomplished by rescaling everywhere the variable $x$ to the form $x/L$, where $L= \pi /2$.    The outcome is presented in the comparative Fig. 9.

To analyze spectral affinities between ${\hat{H}}$ of Eqs. (10-(3) and  the rectangular double well Hamiltonian (c.f. Eq. (27)), additional precautions need to be observed.  The original numerical evaluation of up to eight  eigenvalues and eigenfunctions  involves what we have identified as $\hat{H}_{rect} = -{\frac{1}2} \Delta + V (x)$, with the detailed definition of the rectangular double-well potential given in Section III.B.

To compare  these results  with the $[-1,1]$, $D=1$ setting of  $\hat{H}$, as visualized in Figs. (6) and (7), all numerically obtained data (we have employed Mathematica 12 routines, \cite{blinder}) must ultimately be converted  from the original $[0,\pi ]$, $D=1/2$ framework  to the superharmonic one.

We show how our procedure works, by means of the  exemplary data set in  Table I.  We emphasize that the spectral solution is sought for  $\hat{H}_{well}$, and from the outset we are interested in the nonnegative spectrum, including the bottom   eigenvalue  $E=V_0$, or a close neraby candidate value (this computation is quite sensitive to a proper choice of $a$ and $V_0$, and basically much more than first four decimal digits are needed to get the   exact result (this is untenable  within Mathematica routines, hence some flexibility must be admitted).

Once a spectral solution  of  $\hat{H}_{rect}$ is numerically retrieved, we must compensate the extra factor $D=1/2$ preceding the Laplacian, by considering $2 \hat{H}_{rect}$. This amounts to the  doubling of all computed eigenvalues.

To enable a comparison with the superharmonic case, we need one more correction, actually a conversion of obtained spectral data to the $[-1,1]$ regime.  This may be accomplished by means of a factor $\pi ^2/4$. In view of the energy doubling, mentioned before, an overall conversion factor reads $2\times (\pi ^2/4)= \pi ^2/2$.

\begin{table*}[h]
\begin{tabular}{|c|c|c|c|c|c|c|}
\hline
numerics, \cite{blinder} & renormalization &  conversion   & relabeling   \\ \hline
$E_1= 41,8 =V_0 $ &$ E_1-V_0 =0$  &  $\times  \pi ^2/2 =0$ &   $\rightarrow $   $E_0$   \\ \hline
$E_2=42,4 $ &$E_2-V_0= 0,6$ & $2,96$ & $\rightarrow   E_1$    \\ \hline
$E_3= 44,10$ & $E_3-V_0= 2,3$ & $11.35$ & $\rightarrow E_2$   \\ \hline
$E_4= 46,9$  &$E_4-V_0= 5,1$ & $25,167$ & $\rightarrow E_3$  \\ \hline
$E_5 = 50,8$  & $E_5-V_0= 9,0$ & $44,41$ &  $\rightarrow E_4$   \\ \hline
$E_6= 55,8$ &$ E_6-V_0=14,0 $&  $69,087$  & $\rightarrow E_5$  \\ \hline
\end{tabular}
\caption{Exemplary  data conversion table  for   Mathematica 12 spectral  outcomes  of  the rectangular double well, \cite{blinder}. Input data: overall  well width  $[0,\pi ]$, height of the barrier $V_0= 41,8$, barrier width $a= 2,96$, the  multiplicative  conversion factor for the spectrum is  $\pi ^2/2 = 4,9348$.  Output data:  overall  well width $[-1,1]$, depth of each  well is $- V_0$,  the  "plateau" $V(x)=0$   width contraction factor is $2/\pi $ (takes $L= \pi $ into $L=2$). Hence the  "plateau" size is $2,96 \times  (2/\pi ) = 1,8754 $.}
\end{table*}

\begin{figure}[h]
\begin{center}
\centering
\includegraphics[width=0.6\columnwidth]{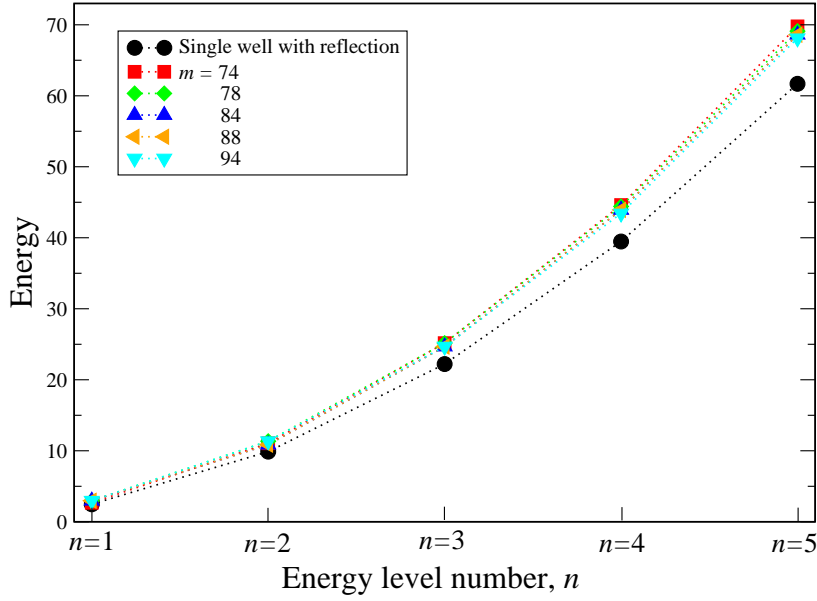}
\caption{We depict comparatively  positive  eigenvalues of the "renormalized"  operator   $\hat{H}_{ren}$, Eq. (28),   on [-1,1], while set in correspondence with the superharmonic $\hat{H}$ for $\kappa_m=m^2$, and  these corresponding to the standard Neumann spectral problem on $[-1,1]$  (here named "single well with reflection").  In addition to  graphical comparisons, in Table II we  collect
numerically and analytically  computed  eigenvalues,  in their  explicit form.}
\end{center}
\end{figure}

In below, in Table II  we reproduce  five   lowest {\it  positive}  eigenvalues $E_n (m), i=1, 2, 3, 4, 5$ of  the   "renormalized"  rectangular double well energy operators
\be
\hat{H}_{ren}= - \Delta  + 2[V(x) - V_0],
\ee
while set  on [-1,1], provided  $V(x)$  is the   best  approximate  fit to   $m=74, 78, 84, 88, 94$   superharmonic potential ${\cal{V}}(x)$,  $\kappa _m=m^2$.   These numerical outcomes   are  set in comparison  with  positive  eigenvalues of   the Neumann  operator  $-\Delta _{\cal{N}}$ on $[-1,1]$, $E_n =(\pi ^2/4) n^2 $.

We point out that  $E_n(m)$   outcomes of    Mathematica 12 routines, \cite{blinder},  for the rectangular double-well, Table II,   originally refer to    the interval $[0, \pi ]$,   the diffusion coefficient  $D=1/2$  and the barrier height $V_0(m)$.  The reproduced data   follow from  the  $(\pi ^2/2) [E_n(m) - V_0(m)]$   conversion recipe, where the  data-converting  coefficient $\pi ^2/2$ adjusts the original rectangular well  data to the $[-1,1]$,  $D=1$ setting.  See  e.g. a complementary   Fig. 11.\\

\begin{table*}[h]
\begin{tabular}{|c|c|c|c|c|c|}
\hline
$n$ & 1 & 2 & 3 & 4 & 5  \\ \hline
well &2.4674 & 9.8696 & 22.2066 & 39.4784& 61.6850   \\ \hline
$m=74$ &2.6647 & 11.054 & 25.16785 &  44.6106 & 69.78    \\ \hline
$m=78$ &2.961 & 11.35 & 25.1675 & 44.4132 &  69.1   \\ \hline
$m=84$  &2.961 & 10.86 & 24.674 & 43.92 &  68.594  \\ \hline
$m=88$  &2.961 & 10.86 & 24.674 & 43.92 & 68.594   \\ \hline
$m=94$ &2.961 &   11.35  & 24.674 & 43.4262 & 68.10  \\ \hline
\end{tabular}
\caption{Lowest {\it  positive}  eigenvalues $E_n (m)$ of  the   "renormalized"  operator   $\hat{H}_{ren}$, Eq. (28),   on [-1,1]  are presented (the eigenavlue zero is thus not displayed).  Rectangular potential contours $V(x)$ provide  best fit  approximations to   $m=74, 78, 84, 88, 94$   superharmonic contours of  ${\cal{V}}(x)$, with $\kappa _m=m^2$.  For comparison we have included   positive  eigenvalues   ($E_0 \sim 0$ being implicit) of   the Neumann  operator  $-\Delta _{\cal{N}}$ on $[-1,1]$, $E_n =(\pi ^2/4) n^2 $. Note: the  eigenvalues were computed for graphical representation purposes, hence their resolution is not high. Even, if appearing as  identical, they  actually  might  differ in higher decimal digits. }
\end{table*}

{\it Remark 1:}  Since we have invoked the standard Neumann well notion, let us  briefly  comment on this, \cite{jph}, \cite{linetsky}-\cite{pilipenko}.  We use the term Neumann Laplacian for the standard Laplacian, while restricted to the interval on $R$ and subject to reflecting (e.g. Neumann) boundary conditions.
In this case, any solution of the diffusion equation
\be
\partial _t \Psi (x,t) = \Delta _{\cal{ N}} \Psi (x,t)
\ee
while restricted to $[-L,L]\subset R$, $L>0$  of length $2L$,  needs to respect  the boundary data
\be
(\partial _x\Psi)(-L,t)= 0 = (\partial _x\Psi)(+L,t)
\ee
for all $t\geq 0$.    The solution of the   Neumann  spectral problem  $ -\Delta _{\cal{ N}} \psi (x) = E \psi (x)$ comprises  eigenfunctions  $\{1/\sqrt{2L}, (1/\sqrt{L})\,\cos[(n\pi /2L)(x+L);\,  n=1,2,3,...\}$ and eigenvalues $ \{ E_0=0, E_1= (\pi/2L)^2,..., E_n= n^2 E_1;\,  n=1,2,3,...\}$.  The choice of $L=1$ maps the problem to the interval $[-1,1]$.\\

{\it Remark 2:}  One should realize that more familiar Dirichlet  spectral  problem  $- \Delta_{\cal{D}} \psi (x) = E\psi (x)$, in the interval $[-L,L]$, typically involves the  boundary  conditions $\psi (-L)=0  = \psi (L)$. As a consequence the spectrum is strictly positive, with $\{E_n= n^2 (\pi /2L)^2; \,  n=1,2,...\}$, while the eigenfunctions have the form   $ (1/\sqrt{L})\,   \{ \sin [(n\pi /2L) (x+L)];\,  n=1,2,3,...\}$.\\

\subsection{Numerical experimentation  with the barrier width vs height  options: quantifying  the jeopardies.}

We take advantage of the existence of the dedicated  Mathematica  routine, \cite{blinder}, which has been  tailored to  yield an "exact solution for rectangular double-well potential". Actually,  up to eight lowest eigenvalues and eigenfunctions of the operator $\hat{H}_{rect} = -{\frac{1}2} \Delta + V (x)$ ,   can be  numerically  retrieved,   with a moderate accuracy,  whose efficiency testing is possible due to (i) fine-tuning  options of  the   barrier  width-height balance,  and (ii)  the presumed validity of Eqs. (25) and (26).

At this point  we recall that  our intention is to get an approximate  ground state function for the superhamonic double-well problem  corresponding to $\kappa_m=m^2$, with the choice of $m$ ranging from  $m=74$   to $94$ (see Table I),  and possibly higher values of $m$.  The problem is that up to  $m= 85$, c.f. Fig. 9,  in the fitting procedure  we may encounter potentially dangerous deviations from  $\{a, V_0\}$ values that guarantee the existence of the sharp eigenvalue zero in the best-fit rectangular well problem.

We have found that the rectangular  double-well with   size parameters  $a=2.77$  and $V_0=  35,56$, allows for the best-fit  numerical evaluation of the  approximate   ground state function  of the superharmonic double-well Hailtonian $\hat{H}$, in case of  $\kappa _m=m=74$.
We note that the  rectangular well with the above parameters  {\it  does not admit a sharp eigenvalue zero}.  This is impossible, in view of Eqs. (25), (26), and the eigenvalue slightly deviates from zero,  maintaining  an   "almost zero"  status.

 We reproduce in   below, in Fig. 11, the   numerically retrieved bottom eigenvalue and the eigenfunction shape   along the barrier "plateau"   in the  vicinity of the endpoint $1$. The  bottom eigenvalue, which we denote $E_0$, following the conventions of Table I is small indeed, and equals $E_1=  35,6   \to E_1- V_0= 0, 04 \to 0, 1974$.

The ground state takes the  form of  a constant function   $1/\sqrt{2}$, extending  roughly  between $[-0.96, +0.96$.  This behavior needs to be yet reconciled   with the Dirichlet condition  imposed  at $1$, being necessarily secured along the   $2\triangle $  remnant  of the full interval $[-1,1]$, by means of the   sine   function, c.f. Section III.B.

 \begin{figure}[h]
\begin{center}
\centering
\includegraphics[width=0.4\columnwidth]{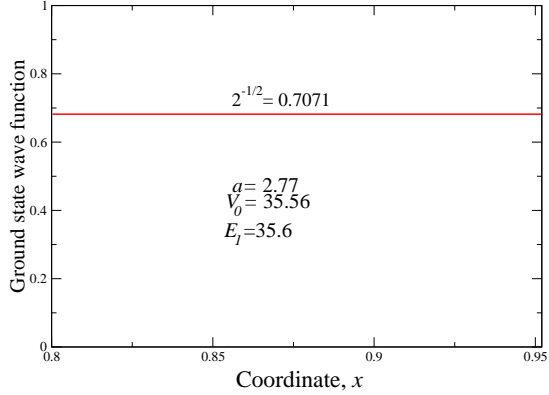}
\caption{An outcome of the Mathematica 12 computation according to Ref. \cite{blinder}. The  rectangular double-well  ground state function is fapp (for all practical purposes) constant and equal $1/\sqrt{2}$ in the interval $[-0,96,+0,96]$.  Notice that  $E-V_0=0.04>0$. Accordingly,  within the employed  numerical accuracy, the  $+0.4$  detuning of $E$  from $V_0$  is insignificant as far as the shape (roughly constant) of the ground state function is concerned. A warning: we  have not covered   the  full  interval [-1,1],  since the  adopted numerical  routine fails in  the  infinitesimal vicinity of the boundaries $\pm 1$.}
\end{center}
\end{figure}

 In the considered case (data of Fig. 11), we deal with the following sequence  of computed eigenvalues  (we   strictly observe conventions of Table I):

\begin{table*}[h]
\begin{tabular}{|c|c|c|}
\hline
$E_1=35,6 \to E_0=0,1974$ & $E_2=36,1 \to E_1=  2,6647 $ & $E_3=37,8 \to E_2= 11.054$ \\ \hline
 $E_4= 40,7 \to E_3=25,1678$& $E_5=44,6 \to E_4=44,61$ & $E_6=49,7 \to E_5=69,78$ \\ \hline
\end{tabular}
\caption{Rectangular double-well input: $a=2,77$, $V_0=35,56$.}
\end{table*}

This  outcome  needs to be  compared with the second, $m=74$,  row  of Table II, where the  eigenvalue (fapp !) $E_0=0$ has been omitted.   Let us note that if we  look  seriously  for the zero energy eigenfunction  with the barrier height measure $V_0= 35,56$, we need to deduce  the corresponding value of $a= a(V_0)$, by using Eq. (25). The result is $a=2,769$, hence encouragingly close to $a=2,77$.

To   quantify  the   technical  jeopardy  of "overshooting" the sharp eigenvalue zero by a small number,   with obvious consequences for other  computed  eigenvalues, let us  consider  the reference data: $a=2,77$ and $V_0=36$. These  strictly  comply with the formulas (25), (26),  and thus secure the existence of the eigenvalue zero for the "renormalized" rectangular well Hamiltonian $\hat{H}_{ren}$, Eq. (28). Mathematica  computation  outcomes, while transformed according to the  Table I give rise to (c.f. also Tables II and III):
\begin{table*}[h]
\begin{tabular}{|c|c|c|}
\hline
$E_1=36 \to E_0=0 $ & $E_2=36,5 \to E_1=  2,4674 $ & $E_3=38,2 \to E_2= 10.8565$ \\ \hline
 $E_4= 41 \to E_3= 24,674$& $E_5=45 \to E_4=44,4132$ & $E_6= 50,1 \to E_5=69,58$ \\ \hline
\end{tabular}
\caption{Rectangular double-well input: $a=2,77$, $V_0=36$.}
\end{table*}

Let us indicate  examples of  rectangular double-well data, which imply the eigenvalue zero, and have the width parameter $a$  close to $a=2,77$. Exemplary cases read: $\{ V_0=36,2,\,    a=2,7724\}, \{ V_0=36,4, \, a=2,7734\}, \{ V_0=35,6, \, a=2,7693\}$.  The fine tuning accuracy in Ref. \cite{blinder} is up to two decimal digits.

\subsection{The concept of Neumann cut: Enforcing  Neumann conditions at endpoints of the barrier.}

We have mentioned before that the eigenvalue  solution of the rectangular double-well problem, Section III.B,   for $E\geq V_0$  involves the cosine function  ($\cosh $ refers to tunneling solutions with $E\leq V_0$)  within the  internal  barrier area. The continuity conditions at the barrier endpoints,  connect derivatives of the cosine with sine  tails, c.f. Eqs. (16) and (17), extending  between the  barrier   and   endpoints of the rectangular well  support (i.e. either $\pm 1$ or $0$, $\pi $).

By examining Figs. 5, 6 and 7 (see  e.g. also \cite{jph} for a thorough discussion of the $\kappa _m= 1$ case), we  realize that although the Neumann condition is not realised at the  internal  barrier endpoints, it is worthwhile  to   consider   (comparatively)  a slight modification of the current best-fit  procedure.

Namely, in addition to the   standard  Neumann well  on $[-1,1]$, let us consider a {\it  cut-ff Neumann well}, whose support is contracted from $[-1,1]$, to  $[-1 + 2\triangle , 1- 2\triangle ]$,  with  $\triangle = \triangle (m)$ evaluated for control values of $m=74, 78, 84, 88, 94$, for  each predefined fitting procedure (location of superharmonic well minima and their distance from the interval  $[-1,1]$  endpoints).   First, for the    case of  $\kappa_m= m^2$,   and next for   $\kappa_m = m$.

So  introduced   narrowing  of the  original  support  $[-1,1]$  by  $4\triangle $ (i.e. twice $2\triangle $) cut-off  at  the interval endpoints, defines the new  re-sized Neumann well, which we  call  the  {\it Neumann cut}.

\begin{figure}[h]
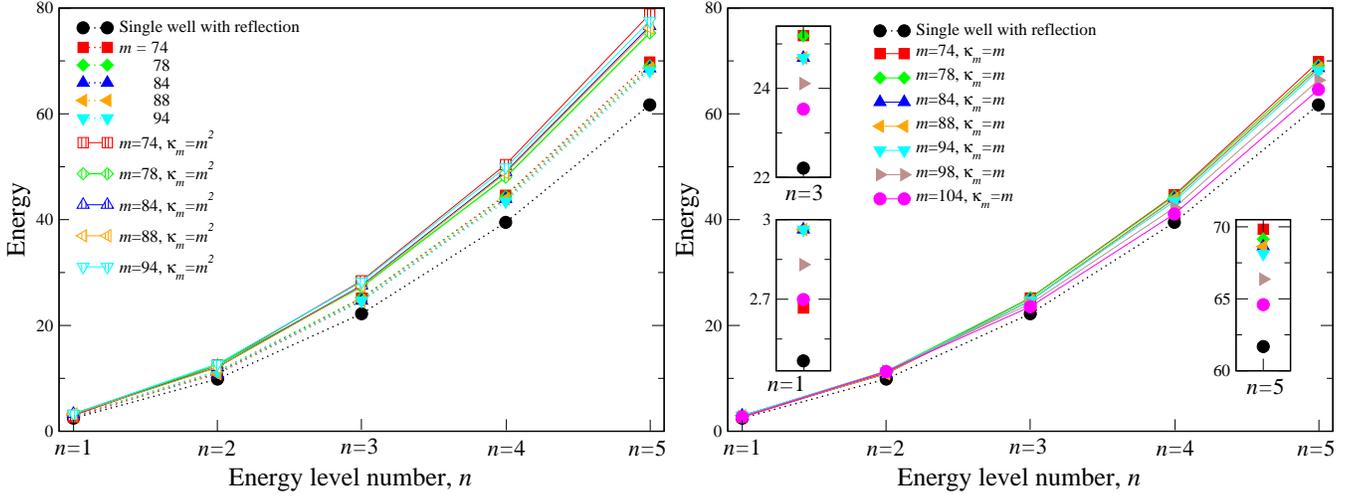

\begin{center}
\centering
\includegraphics[width=0.49\columnwidth]{widmo3a}
\includegraphics[width=0.49\columnwidth]{widmo3d}
\caption{Left panel: we have filtered the spectral  data to depict comparatively, (i) Neumann cut  for  $\kappa_m=m^2$ (top), (ii)  best-fit   rectangular well approximation of the superharmonic double-well for  $\kappa _m=m^2 $ (middle), (iii)  Neumann well spectrum  for $n>0$ (bottom).  Right panel: A comparative display of, (i) Neumann cut for $\kappa_m=m$  (top data), (ii)   Neumann well  (single well with reflection) spectrum (bottom).  Insets display a location of numerically retireved  eigenvalues, up to  $m=104 $.}
\end{center}
\end{figure}

\begin{table*}[h]
\begin{tabular}{|c|c|c|c|c|c|c|c|}
\hline
$m=94$ & N-cut, $\kappa _m=m^2$   &  N-cut, $\kappa_m=m $ & rect.  well, $\kappa_m = m^2 $ & N-cut, $\kappa_m =m=98$ &   N-cut, $\kappa_m=m =104$  &   Neumann well    \\ \hline
$E_1$ & $3.273932$    & $2.962383  $   & $>\,  2.961\, >$ &  $2.830433  $   & $2.698934$  & $  2.4674 $   \\ \hline
$E_2 $ &$12.549520$    &$11.3552029  $ & $ > \,   11.35\,  >$ & $11.3253029 $     & $11.294264$  & $ 9.8696  $    \\ \hline
$E_3 $ & $27.281663$     & $24.685528 $  & $ > \,  24.674\, > $&  $24.109515 $  &$23.533541$    & $22.2066   $   \\ \hline
$E_4$  &$48.015727$   &$  43.446529  $& $  > \, 43.4262\,  >$ &  $42.254529 $ & $41.062529$      & $ 39.4784  $  \\ \hline
$E_5$ & $75.297125$   & $ 68.131817  $ & $ > \, 68.10\,  >$ & $ 66.361801 $ & $64.591806$   & $ 61.850 $   \\ \hline
\end{tabular}
\caption{We depict  positive ($E_0=0$ being kept in memory) eigenvalue data  for the choice  of   $m=94$, for the following  computation regimes: (i) Neumann cut (denoted N-cut) in the   rectangular well approximation of the  superharmonic double-well,  $\kappa_m=m^2$; (ii) Neumann cut  for $\kappa _m =m $, (iii)   best-fit   rectangular well approximation of the superharmonic double-well for  $\kappa =m^2 $,  c.f. Table II and Fig. 10 (inequality symbols indicate lower and upper estimates provided respectively by the data (iv) and (ii));  (iv) standard  Neumann well spectrum  for $n>0$.   Additionally, we have depicted two   columns of data  corresponding  to $m=98$ and  $m=104$, for  the  $\kappa_m=m$   Neumann cut.}
\end{table*}

For the record, we mention that the $m=104$   data,   reported  in Fig. 12 and Table IV,  have been obtained through averaging over 15 repeated computation  runs, with   somewhat  diverse outcomes. Effectively, the case of $m=104$ stays on the  verge of Ref. \cite{blinder}    computing  capabilities.

\section{Conclusions and contexts.}

The major observation coming from our discussion  in Section IV,  stems from  spectral  data reported in  Fig. 12 and  Table IV.  We have demonstrated that  a numerical evaluation of lowest eigenvalues  in the   rectangular double-well approximation of the  superharmonic double-well, for $\kappa_m=m^2$, $m=94$ is possible. The corresponding eigenfunctions (not depicted in the present paper) are retrievable as well.  This in turn gives meaning to the spectral relaxation scenario of the original Smoluchowski process, in the least up to $m=104$.

The numerically retrieved  spectral  outcome  can be effectively controlled  and justified  by two-sided estimates  set by   {\it  exact } spectral  solutions of two resized Neumann wells (Neumann cuts with $\kappa_m=m$). The pertinent Neumann cuts are  set  sharply  upon  interior barriers of  rectangular double-well approximants, and effectively involve the validity on the Neumann condition  at endpoints of resized support intervals.  We point out that the Neumann cuts  correspond  to: (i) $\kappa_m=m=94$   (upper bound) and   (ii) $\kappa_m=m=98$   (lower bound).

Our approximate solvability argument for the spectral problem of the superharmonic system   $\hat{H}$  directly employs the "renormalized" rectangular double well system $\hat{H}_{ren} = -(1/2)\Delta + [V(x) - V_0]$, where $V_0$ is the height of the double-well  barrier ($-V_0$  measures the depth of local wells in the corresponding superharmonic double-well system). Therefore  we are convinced that the  approximation validity, as $m$ grows indefinitely, becomes questionable    both on the formal and physical grounds. Our computation procedure (modulo the numerical  programming  adjustments) is operational for each finite value of $m <\infty $.

The presented analysis of a particular spectral problem  for a superharmonic double-well  Hamiltonian  $\hat{H}$, Eq. (1),  has been motivated by the method  of eigenfunction expansions, often used in the analysis of spectral relaxation of diffusion processes, \cite{risken,pavl,jph}. Somewhat surprisingly, the technical difficulty in solving this class of spectral problems is not exceptional, and is shared by a broad class of so called "quasi-exactly solvable"  Schr\"{o}dinger systems, \cite{turbiner,turbiner1}, see also \cite{baner,brandon,maiz,okopinska}. "Quasi-solvability" is here a misnomer, because the   solvability of the  pertinent systems is not excluded, but extremally limited to some special cases.

The systems studied in Ref. \cite{turbiner} are most easily constructed by means of the method employed in the present paper, where basically  any  positive  $L^2(R)$-normalized function  may serve as a square root of a certain probability distribution on $R$.  A variety of Hamiltonian systems, with potentials of the generic form ${\cal{V}}(x)= [\Delta \rho _*^{1/2}(x)]/\rho _*^{1/2}(x)$  can  be (re)constructed this way. In particular, the same route has been followed in Refs. \cite{streit,zambrini,vilela,faris,jph,jph1}, while guided by the idea of "reconstruction of  (random) dynamics from the eigenstate".

A peculiarity of all mentioned Schr\"{o}dinger-type systems is that their ground state , by construction has been associated with the zero binding  energy. Nonetheless, an issue of zero energy ground states is not anything  close to being exotic. One may mention fairly serious research on bound states embedded in the continuum, and  bound states with  zero   energy, \cite{lorinczi,lorinczi1,lorinczi2}.  Less mathematically advanced research concerning zero energy bound states can also be mentioned, \cite{nieto,makowski,robinett}. This in line with a complementary resarch on zero-curvature eigenstates, \cite{ahmed,gilbert1}.

The general issue of boundary conditions in case of impenetrable barriers, complementary to \cite{jph,carlsaw}, has been addressed in \cite{robinett,karw,diaz}.

We point out that  the  Langevin-induced Fokker-Planck equations have been  solved for potentials of the rectangular double-well shape, following \cite{risken,kampen} and \cite{kostin,risken1,so}. It might be of interest to investigate comparatively, Langevin-Fokker-Planck problems with drifts stemming (through negative gradients) from superharmonic double well potentials (3).

\end{document}